\documentclass[preprint,showpacs,amsmath,amssymb,aps,prd]{revtex4}

\usepackage{graphicx}

\usepackage{amsmath}

 \newcommand{\ds}{\displaystyle}

\begin{document}

\title{Searching for MOND in scalar-tensor theories of gravity}

\author{Milovan Vasili\'c} \email{mvasilic@ipb.ac.rs}
\affiliation{Institute of Physics, University of Belgrade, P.O.Box 57,
11001 Belgrade, Serbia}

\date{\today}

\begin{abstract}
In this paper, I study spherically symmetric solutions in a simple
class of geometric sigma models of the Universe. This class of models
is a subclass of the wider class of scalar-tensor theories of gravity.
The purpose of this work is to examine how the additional scalar degree of
freedom modifies Newtonian gravitational force. The general solution for
spherically symmetric metric far from the point source is obtained in
a weak field approximation. As it turns out, it is parametrized by the
mass of the source, and an additional function of time. One particular
model is examined as an example. It is shown that there are solutions
that accommodate MOND regime at some distances from the source.
Unfortunately, the obtained interval of distances turns out to be
smaller than it is needed. An additional analysis shows that genuine
MOND, that explains galactic curves of all nearby galaxies, cannot be
obtained.
\end{abstract}


\maketitle

\section{Introduction}\label{Sec1}

With this work, I begin the search for a plausible explanation of the
observed galactic curves. While Newtonian gravity successfully
explains trajectories of planets in the Solar system, and trajectories
of stars close to the galaxy center, it encounters problems when
trying to explain trajectories of more distant stars. Indeed, the
observed trajectories of stars turn out to require more galactic mass
than our telescopes can detect \cite{1,2,3,4,5,6,7,8}. The missing
mass is commonly referred to as {\it dark matter}. With the presence
of dark matter, the explanation of flat galactic curves is quite
simple. Still, there are two problems that cannot be ignored. The
first is that distribution of dark matter in every galaxy must be fine
tuned to produce the observed galactic curves. Then, one needs an
explanation of how this specific distribution of mass has been formed
in the first place. The second is the very nature of dark matter. We
do not know if it is some sort of exotic matter, or it is just
ordinary matter which is, for some reason, not seen. Another
explanation of flat galactic curves is that Newtonian gravity is
somehow modified at large distances. One of the most successful
phenomenological models is Milgrom's modified Newtonian dynamics,
commonly referred to as MOND \cite{9,10,10a,11,12,13,14,15}. This
model does not require dark matter, but it suffers from the
nonexistence of a satisfactory underlying theory. In this work, I
shall search for MOND in a class of scalar-tensor theories of gravity.

The motivation for the study of scalar-tensor theories comes from the
fact that these have already been widely used as simplest inflationary
cosmological models. They complement the standard cosmological model
($\Lambda$CDM) which, although very successful in explaining the
present epoch, lacks inflation. Scalar-tensor theories belong to the
class of modified theories of gravity. A logical step is to check if
this modification has anything to do with MOND.

In what follows, I shall examine a class of geometric sigma models
with one scalar field. These are a subclass of scalar-tensor theories
of gravity, and differ from ordinary sigma models in two respects.
First, their scalar fields can all be gauged away, leaving us with the
metric alone. Second, they are constructed in an unconventional way.
One first chooses metric one would like to be the vacuum of the model,
and then builds a theory that has this metric as its solution. These
models have first been proposed in Ref.\ \cite{16} in the context of
fermionic excitations of flat geometry. In Refs.\ \cite{17} and
\cite{18}, they are used for the construction of various inflationary
and bouncing cosmologies. The purpose of this work is to examine how
the additional scalar degree of freedom modifies Newtonian
gravitational force. To this end, I shall consider a pointlike source,
and calculate spherically symmetric metric far from it.

The results of the paper are summarized as follows. First, I obtained
general spherically symmetric solution to the whole class of
considered geometric sigma models. The solution is obtained in a weak
field approximation, and in the form of a power series. As it turns
out, this general solution has two free parameters. The first is a
constant with the dimension of mass, and the second is an arbitrary
function of time. The presence of a free function of time in the
general solution is caused by the additional scalar degree of freedom.
The second result concerns the comparison of the obtained solution
with MOND. A specific example has been analyzed. I have shown that
there exists a particular class of solutions that mimic MOND in a
finite interval of distances, and a wide range of source masses.
Sadly, the obtained interval of distances turns out to be much smaller
than the original MOND suggests. To check if the second result is a
general property of the model, I have done a nonperturbative analysis
of field equations. The result is that exact MOND can always be
obtained at a fixed moment of time, such as the present epoch, but its
time evolution is so rapid that MOND regime is almost instantly lost.
In particular, it is seen that even the nearest galaxies are devoid of
the exact MOND.

The layout of the paper is as follows. In Sec.\ \ref{Sec2}, a precise
definition of the class of models to be considered is given. The very
construction of geometric sigma models is only briefly recapitulated.
In Sec.\ \ref{Sec3}, spherically symmetric ansatz is applied to 
field equations. The general solution is obtained in a weak field
approximation, and in the form of a power series. Two free parameters
parametrize the solution, and one of them is a free function of time.
In Sec.\ \ref{Sec4}, a particular geometric sigma model is examined for
a class of spherically symmetric solutions. This class is defined by
the simplest nontrivial choice of the time dependent parameter. MOND
behavior is found in a finite interval of distances, and for a wide
range of source masses. In Sec.\ \ref{Sec5}, a nonperturbative analysis
of field equations is done. It is shown that, although MOND can always
be obtained at a fixed moment of time, its time evolution is so rapid
that it almost instantly disappears. Sec.\ \ref{Sec6} is devoted to
concluding remarks.

My conventions are as follows. Indexes $\mu$, $\nu$, ... and $i$, $j$,
... from the middle of alphabet take values $0,1,2,3$. Indexes
$\alpha$, $\beta$, ... and $a$, $b$, ... from the beginning of
alphabet take values $1,2,3$. Spacetime coordinates are denoted by
$x^{\mu}$, ordinary differentiation uses comma ($X_{,\,\mu} \equiv
\partial_{\mu} X$), and covariant differentiation uses semicolon
($X_{;\,\mu}\equiv \nabla_{\mu}X$). Repeated indexes denote summation:
$X_{\alpha\alpha} \equiv X_{11} + X_{25} + X_{33}$. Signature of the
$4$-metric $g_{\mu\nu}$ is $(-,+,+,+)$, and curvature tensor is
defined as $R^{\mu}{}_{\nu\lambda\rho} \equiv \partial_{\lambda}
\Gamma^{\mu}{}_{\nu\rho} - \partial_{\rho}
\Gamma^{\mu}{}_{\nu\lambda}+\Gamma^{\mu}{}_{\sigma\lambda}
\Gamma^{\sigma}{}_{\nu\rho} - \Gamma^{\mu}{}_{\sigma\rho}
\Gamma^{\sigma}{}_{\nu\lambda}$. Throughout the paper, the natural
units $c=\hbar=1$ are used.

\section{Model}\label{Sec2}

The model considered in this paper belongs to the class of geometric
sigma models, originally defined in Ref.\ \cite{16}. The main feature
of every geometric sigma model is that it is defined by associating
action functional with a fixed freely chosen metric
$g_{\mu\nu}^{(o)}(x)$. The action has the form
\begin{equation} \label{1} 
I_g = \frac{1}{2\kappa}\int d^4x\sqrt{-g}\left[ R -
F_{ij}(\phi)\phi^i_{,\mu}\phi^{j,\mu} - V(\phi) \right] ,
\end{equation}
where $F_{ij}(\phi)$ and $V(\phi)$ are target metric and potential of
four scalar fields $\phi^i(x)$. The constant $\kappa \equiv 8\pi G$
stands for the gravitational coupling constant. The target metric
$F_{ij}(\phi)$ is constructed by replacing $x^i$ with $\phi^i$ in the
expression
\begin{equation} \label{2} 
F_{ij}(x)\equiv R_{ij}^{(o)}(x) - \frac{1}{2} V(x) g_{ij}^{(o)}(x) \,, 
\end{equation}
where $R_{\mu\nu}^{(o)}(x)$ is Ricci tensor for the metric
$g_{\mu\nu}^{(o)}(x)$. The same replacement in an arbitrary function
$V(x)$ defines the potential $V(\phi)$. This construction guarantees that
\begin{equation} \label{3} 
\phi^i = x^i \,, \quad g_{\mu\nu} = g_{\mu\nu}^{(o)} 
\end{equation} 
is a solution of the field equations defined by Eq.\ (\ref{1}). In
what follows, the solution Eq.\ (\ref{3}) will be referred to as {\it
vacuum}. It is seen that physics of small perturbations of this vacuum
allows the gauge condition $\phi^i(x) = x^i$. Gauge fixed field
equations depend on the metric alone.

In cosmology, the vacuum metric $g_{\mu\nu}^{(o)}$ is chosen to be
the background metric
\begin{equation} \label{4} 
ds^2 = -dt^2 + a^2(t)\left( dx^2 + dy^2 + dz^2 \right). 
\end{equation} 
The scale factor $a(t)$ is an unspecified function of time, and it
will remain unspecified throughout most of the paper. In other words,
I am looking for the general spherically symmetric solution in an
arbitrary background. The model is defined by determining $V(x)$ and
$F_{ij}(x)$. While $V(x)$ is kept arbitrary, $F_{ij}(x)$ is determined by
Eq.\ (\ref{2}). One obtains
\begin{equation}\label{5} 
F_{00} = W - 2\dot H \,, \quad F_{0b} = 0 \,,\quad
F_{ab}=-a^2\,W \delta_{ab} \,, 
\end{equation} 
where $H\equiv \dot a/a$ is the Hubble parameter, and $W$ is defined
by
\begin{equation} \label{6} 
V \equiv 2\left( W + \dot H + 3H^2 \right) . 
\end{equation} 
The ``dot'' denotes time derivative. The target metric $F_{ij}(\phi)$
and the potential $V(\phi)$ are obtained by the substitution $x^i \to
\phi^i$ in $F_{ij}(x)$ and $V(x)$.

In what follows, I shall consider the simplest case $W=0$.
Then, the target metric becomes degenerate, as $F_{00}$ remains the
only nonzero component. Owing to its independence of spatial
coordinates, the resulting action depends on one scalar field only.
Precisely,
\begin{equation} \label{7} 
I_g = \frac{1}{2\kappa}\int d^4x\sqrt{-g}\left[ R -
F(\phi)\phi_{,\mu}\phi^{,\mu} - V(\phi) \right] ,
\end{equation}
where $F \equiv F_{00}$ and $\phi \equiv \phi^0$. The action Eq.\
(\ref{7}) defines a class of scalar-tensor theories parametrized by
the scale factor $a(t)$. The target metric and potential are defined
by the substitution $t \to \phi$ in
\begin{equation} \label{8}
F(t) \equiv -2\dot H \,, \quad 
V(t) \equiv 2\left( \dot H + 3H^2 \right) . 
\end{equation} 
The scalar $\phi$ is commonly referred to as {\it inflaton}. More
detailed construction of geometric sigma models with one scalar field
can be found in Ref.\ \cite{17}.

\section{Field equations}\label{Sec3}

The complete action needed for considerations of this paper must
include matter fields. It has the form
\begin{equation} \label{9}
I = I_g + I_m \,,
\end{equation} 
where $I_g$ is geometric action Eq.\ (\ref{7}), and $I_m$ stands for
the action of matter fields. The matter Lagrangian is assumed to be
that of the standard model of elementary particles minimally coupled
to gravity. This means that direct matter-inflaton coupling is absent,
which leaves us with the following field equations. First, equation
\begin{equation} \label{10}
R_{\mu\nu}-\frac 12 g_{\mu\nu}R =
T^{\phi}_{\mu\nu} + \kappa\, T^{m}_{\mu\nu} \,,
\end{equation} 
is obtained by varying the action Eq.\ (\ref{9}) with respect to the
metric. Tensors on the right-hand side of Eq.\ (\ref{10}) stand for
the stress-energy of the inflaton and matter fields, respectively.
Specifically,
\begin{equation} \label{11}
T^{m}_{\mu\nu} = - \frac{2}{\sqrt{-g}}\,
\frac{\delta I_m}{\delta g^{\mu\nu}} \,,
\end{equation}
\begin{equation} \label{12}
T^{\phi}_{\mu\nu} = G_{\mu\nu}-\frac 12 
\left( G^{\rho}_{\,\rho}+V \right) g_{\mu\nu} \,,
\end{equation} 
where
$$
G_{\mu\nu} \equiv F(\phi)\, \phi_{,\mu}\phi_{,\nu} \,. 
$$
The inflaton field equation is obtained by varying the action Eq.\
(\ref{9}) with respect to $\phi$. Owing to the absence of direct
matter-inflaton coupling, the same equation is obtained by varying
geometric action Eq.\ (\ref{7}). Thus obtained inflaton equation
implies covariant conservation of the inflaton stress-energy tensor,
\begin{equation} \label{13}
\nabla^{\mu}T^{\phi}_{\mu\nu} = 0 \,.
\end{equation}
In fact, the inflaton equation is equivalent to Eq.\ (\ref{13}), as
only one out of these four equations is truly independent.
Finally, matter field equations are obtained by varying the action
Eq.\ (\ref{9}) with respect to matter fields. Specific form of these
equations is not known unless $I_m$ is specified. However, some
information on the dynamics of matter fields can be obtained directly
from Eqs.\ (\ref{10}) and (\ref{13}). Indeed, the Bianchi identities
imply covariant conservation of the right-hand side of Eq.\
(\ref{10}). With the help of Eq.\ (\ref{13}), one then obtains
\begin{equation} \label{14}
\nabla^{\mu}T^{m}_{\mu\nu} = 0 \,.
\end{equation}
Thus, the two stress-energy tensors, $T^{\phi}_{\mu\nu}$ and
$T^{m}_{\mu\nu}$, are independently covariantly conserved.

In what follows, matter fields are assumed to be localized in a point,
which I choose to be $\vec x = 0$. Then, the field equations in the
region $\vec x \neq 0$ reduce to those obtained from the geometric
action Eq.\ (\ref{7}). It is easily checked that they possess the
vacuum solution
$$
\phi = t \,, \quad g_{\mu\nu} = g_{\mu\nu}^{(o)} \,,
$$
where $g_{\mu\nu}^{(o)}$ is defined by Eq.\ (\ref{4}). In this work, I
shall consider spherically symmetric deviations from this vacuum,
caused by the presence of a massive particle in $\vec x = 0$. As these
deviations are expected to be small far from $\vec x = 0$, one is
allowed to fix the gauge $\phi = t$. In this gauge, the inflaton
equation is identically satisfied, and we are left with Eq.\ (\ref{10}).
Using the definition Eq.\ (\ref{2}), it is rewritten as
\begin{equation} \label{15}
R_{\mu\nu} - R_{\mu\nu}^{(o)} - \frac{V}{2} 
\left( g_{\mu\nu} - g_{\mu\nu}^{(o)} \right) = 0 \,.
\end{equation}
This noncovariant equation carries the full content of the model in
the gauge $\phi =t$. The residual coordinate transformations are
$$
x^{\alpha} \to x^{\alpha} + \xi^{\alpha}(x) \,,\quad t \to t \,.
$$
They allow for an additional gauge fixing, which I choose to be
$g_{0\alpha}=0$. The adopted gauge fixing conditions
\begin{equation} \label{16}
\phi =t \,, \quad g_{0\alpha}=0
\end{equation}
leave us with time independent parameters $\xi^{\alpha} =
\xi^{\alpha}(\vec x)$.

In what follows, I shall search for spherically symmetric solutions of
Eq.\ (\ref{15}). The most general spherically symmetric metric in the
gauge Eq.\ (\ref{16}) has the form
\begin{equation} \label{17}
g_{00} = \mu \,, \quad
g_{0\alpha} =0 \,, \quad 
g_{\alpha\beta} = \nu P^{\parallel}_{\alpha\beta} +
\rho P^{\perp}_{\alpha\beta} \,,
\end{equation}
where $P^{\parallel}_{\alpha\beta}$ and $P^{\perp}_{\alpha\beta}$ are
parallel and orthogonal projectors on $\vec x$,
\begin{equation} \label{18}
P^{\parallel}_{\alpha\beta} \equiv 
\frac{x^{\alpha}x^{\beta}}{r^2}  \,,   \quad
P^{\perp}_{\alpha\beta} \equiv \delta_{\alpha\beta} -
\frac{x^{\alpha}x^{\beta}}{r^2} \,,
\end{equation}
and $\mu$, $\nu$, $\rho$ are functions of $r$ and $t$, only. The
radius $r$ is defined by $r^2 \equiv x^2+y^2+z^2$. Now, the field
equations (\ref{15}) are straightforwardly expressed in terms of
$\mu$, $\nu$, $\rho$ and the scale factor $a$. For this, one uses Eq.\
(\ref{17}) to calculate $R_{\mu\nu}$, and the vacuum metric Eq.\
(\ref{4}) to determine $R_{\mu\nu}^{(o)}$. The potential $V$ is
defined by Eq.\ (\ref{8}). Skipping the details of cumbersome
calculations, here I display the final result. The `00' component of
Eq.\ (\ref{15}) takes the form
\begin{subequations}\label{19}
\begin{equation}\label{19a}
\begin{array}{rl}
& \ds \frac 14 \bigg[ \frac{\dot\mu}{\mu}\Big(\frac{\dot\nu}{\nu} +
2\frac{\dot\rho}{\rho}\Big) - \Big(\frac{\dot\nu^2}{\nu^2} +
2\frac{\dot\rho^2}{\rho^2}\Big) - 2\Big(\frac{\dot\nu}{\nu} +
2\frac{\dot\rho}{\rho}\Big)_{,0} +                                              \\
& \ds \frac{\mu'}{\nu}\Big( \frac{\mu'}{\mu} - 
\frac{\nu'}{\nu} - 2\frac{\rho'}{\rho}\Big)\bigg] - 
\frac 12 \bigg[\Big(\frac{\mu'}{\nu}\Big)' +
\frac{2}{r}\frac{\mu'}{\nu}\bigg] -                                              \\
& \ds \big(\dot H + 3H^2\big) \mu + 2\dot H = 0 \,, 
\end{array}
\end{equation}
and the three `$0\alpha$' components are all equivalent to the equation
\begin{equation}\label{19b}
\frac 12 \Big(\frac{\dot\nu}{\nu} - \frac{\dot\rho}{\rho}\Big)
\Big(\frac{\rho'}{\rho} + \frac{2}{r}\Big) +
\frac 12 \frac{\mu'}{\mu}\frac{\dot\rho}{\rho} -
\Big(\frac{\dot\rho}{\rho}\Big)' = 0 \,.
\end{equation}
The `$\alpha\beta$' components of Eq.\ (\ref{15}) are shown to
have the form $A P^{\parallel}_{\alpha\beta} + B
P^{\perp}_{\alpha\beta}=0$. They are equivalent to two equations
\begin{equation}\label{19c}
\begin{array}{rl}
&\ds \frac 12 \Big(\frac{\dot\nu}{\mu}\Big)_{,0} + 
\frac 14 \frac{\dot\nu}{\mu}\Big(\frac{\dot\mu}{\mu} -
\frac{\dot\nu}{\nu} + 2\frac{\dot\rho}{\rho}\Big) +
\frac 12 \Big(\frac{\mu'}{\mu} + 2\frac{\rho'}{\rho}\Big)' -           \\
&\ds \frac 14 \frac{\nu'}{\nu}\Big(\frac{\mu'}{\mu} + 
2\frac{\rho'}{\rho}\Big) + \frac 14 \bigg[\Big(\frac{\mu'}{\mu}
\Big)^2 + 2\Big(\frac{\rho'}{\rho}\Big)^2\bigg] -                         \\ 
&\ds \frac{1}{r}\Big(\frac{\nu'}{\nu}-2\frac{\rho'}{\rho}\Big)+ 
\big(\dot H + 3H^2\big) \rho = 0 \,,  
\end{array}
\end{equation}
\begin{equation}\label{19d}
\begin{array}{rl}
&\ds \frac 12 \Big(\frac{\dot\rho}{\mu}\Big)_{,0} + 
\frac 14 \frac{\dot\rho}{\mu}\Big(\frac{\dot\mu}{\mu} +
\frac{\dot\nu}{\nu}\Big) - 
\frac 12 \bigg[ \frac{2}{r}\Big(1-\frac{\rho}{\nu} \Big) -
\frac{\rho'}{\nu} \bigg]' -                                                            \\
&\ds \frac{1}{r}\bigg[\frac{2}{r}\Big(1-\frac{\rho}{\nu}\Big) -
\frac{\rho'}{\nu}\bigg] +
\frac 14 \Big(\frac{\mu'}{\mu} + \frac{\nu'}{\nu}\Big)
\Big(\frac{\rho'}{\nu} + \frac{2}{r}\frac{\rho}{\nu}\Big) +         \\ 
&\ds \big(\dot H + 3H^2\big) \rho = 0 \,.  
\end{array}
\end{equation}
\end{subequations}
To check if Eqs.\ (\ref{19}) are correctly derived, I have
solved these equations in three simplest cases. First, it is easily
verified that the vacuum $\mu=-1$, $\nu=\rho=a^2$ is a solution of
Eqs.\ (\ref{19}) for every $a(t)$. Second, the Schwarzschild metric
$$
\rho = 1 \,, \quad
\mu = -\frac{1}{\nu} = -\left(1-\frac{\ell}{r}\right)
$$
is obtained as the general spherically symmetric solution in the flat
background $a(t)=1$. Finally, I examined de Sitter background
$a(t)=e^{\omega t}$. As expected, the Schwarzschild-de Sitter solution
$$
\rho = 1 \,, \quad
\mu=-\frac{1}{\nu}=-\left(1-\frac{\ell}{r}-\omega^2 r^2\right)
$$
is obtained. The integration constant $\ell$ stands for the
Schwarzschild radius of the point source.

The field equations (\ref{19}) remain unchanged by the action of the
residual coordinate transformation $r\to r+\xi(r)$. With respect to
this, the variables of the theory transform as
\begin{equation}\label{20}
\begin{array}{rl}
&\ds \delta_0 \mu = -\xi \mu' \,,                                    \\ [1.3ex]                                   
&\ds \delta_0 \nu = -\xi \nu' - 2\xi' \nu \,,                      \\ [1.3ex]
&\ds \delta_0 \rho = -\xi \rho' - \frac{2}{r} \xi \rho \,.   
\end{array}
\end{equation}
These transformations define the residual gauge symmetry of Eqs.\
(\ref{19}).

In what follows, I shall search for the general solution of Eqs.\
(\ref{19}), in which $a(t)$ remains unspecified. To this end, I shall
consider the region far from the source, where weak field
approximation can be used. This is because the unperturbed Eqs.\
(\ref{19}) are too complicated to be solved without any approximation.
Thus, I define
\begin{equation}\label{21}
\begin{array}{rl}
&\ds \mu \equiv -1+\mu_1 ,                                \\ [1.3ex]
&\ds \nu \equiv a^2 \left(1+\nu_1\right) ,           \\ [1.3ex]
&\ds \rho \equiv a^2 \left(1+\rho_1\right) .
\end{array}
\end{equation}
The new fields $\mu_1$, $\nu_1$, $\rho_1$ are assumed to be small, so
that quadratic and higher order terms can be neglected. After a
lengthy calculation, the linearized field equations are brought to
the form
\begin{subequations}\label{22}
\begin{equation}\label{22a}
\frac{1}{r}\left[\nu_1 - \left(r\rho_1\right)'\right]_{,0} -
H\mu'_1 = {\cal O}_2 \,,
\end{equation}
\begin{equation}\label{22b}
\begin{array}{rl}
&\ds 2a^2 \left[ H\left(\dot\nu_1+2\dot\rho_1\right) +
\left(\dot H+3H^2\right)\mu_1 \right] +                                   \\
&\ds \frac{2}{r}\left[\nu_1-\left(r\rho_1\right)'\right]' +
\frac{2}{r^2}\left[\nu_1-\left(r\rho_1\right)'\right] =
{\cal O}_2 \,,
\end{array}
\end{equation}
\begin{equation}\label{22c}
\begin{array}{rl}
&\ds a^2\left[\left(\nu_1-\rho_1\right)_{,00} +
3H\left(\nu_1-\rho_1\right)_{,0}\right] + \mu''_1 -                  \\
&\ds \frac{1}{r}\left[\mu_1-\nu_1+
\left(r\rho_1\right)'\right]' - \frac{2}{r^2} \left[\nu_1-
\left(r\rho_1\right)'\right] = {\cal O}_2 \,,   
\end{array}
\end{equation}
\begin{equation}\label{22d}
\dot\nu_1 + 2\dot\rho_1 + \dot\mu_1 +
\bigg(6H+\frac{\ddot H}{\dot H}\bigg)\mu_1 = {\cal O}_2 \,,
\end{equation}
\end{subequations}
where ${\cal O}_2$ stands for quadratic and higher order terms in
$\mu_1$, $\nu_1$, $\rho_1$. The residual gauge symmetry of the
linearized theory is the linearized version of Eq.\ (\ref{20}). One
finds
\begin{equation}\label{23}
\begin{array}{rl}
&\ds \delta_0 \mu_1 = {\cal O}_2 \,,                       \\  [1.3ex]
&\ds \delta_0 \nu_1 = - 2\,\xi' + {\cal O}_2 \,,         \\ [1.3ex]
&\ds \delta_0 \rho_1 = - \frac{2}{r}\,\xi + {\cal O}_2 \,.   
\end{array}
\end{equation}
In the next section, I shall search for the general solution of Eqs.\ (\ref{22}).

\section{Solution}\label{Sec4}

The solution of Eqs.\ (\ref{22}) is searched for in the form of a power
series. Specifically, I use the decomposition
\begin{equation}\label{24}
\begin{array}{rl}
&\ds \mu_1 = \sum_{n=0}^{\infty}\alpha_n r^{n-1} ,    \\ [1.3ex]
&\ds \nu_1 = \sum_{n=0}^{\infty}\beta_n r^{n-1} ,      \\ [1.3ex]
&\ds \rho_1 = \sum_{n=0}^{\infty}\gamma_n r^{n-1} ,
\end{array}
\end{equation}
where $\alpha_n(t)$, $\beta_n(t)$, $\gamma_n(t)$ are time dependent
coefficients. The substitution of Eq.\ (\ref{24}) into Eqs.\
(\ref{22}) yields the following set of ordinary differential
equations. Eq.\ (\ref{22a}) becomes
\begin{equation}\label{25}
A_n \equiv 
\dot\beta_n - n\dot\gamma_n - \left(n-1\right) H\alpha_n = 0 
\end{equation}
for all $n\geq 0$. Eq.\ (\ref{22b}) leads to
\begin{subequations}\label{26}
\begin{equation}\label{26a}
\beta_1 - \gamma_1 = 0 \,,
\end{equation}
\begin{equation}\label{26b}
\begin{array}{rl}
B_n \equiv 
&\ds a^2 H \bigg[ \dot\beta_n + 2\dot\gamma_n + 
\bigg(3H+\frac{\dot H}{H}\bigg)\alpha_n \bigg] +       \\
&\ds \left(n+2\right) \big[ \beta_{n+2} - 
\left(n+2\right) \gamma_{n+2} \big] = 0 \,.
\end{array}
\end{equation}
\end{subequations}
From Eq.\ (\ref{22c}), one finds
\begin{subequations}\label{27}
\begin{equation}\label{27a}
\alpha_0 - \beta_0 = 0 \,,
\end{equation}
\begin{equation}\label{27b}
\begin{array}{rl}
C_n \equiv 
&\ds a^2 \left[ \ddot\beta_n - \ddot\gamma_n + 
3H \left( \dot\beta_n - \dot\gamma_n \right) \right] +    \\
&\ds \left(n-1\right) \big[ \beta_{n+2} - 
\left(n+2\right) \gamma_{n+2} \big] +                          \\
&\ds  \left(n-1\right)\left(n+1\right) 
\alpha_{n+2} = 0 \,,
\end{array}
\end{equation}
\end{subequations}
and from Eq.\ (\ref{22d})
\begin{equation}\label{28}
D_n \equiv \dot\beta_n + 2\dot\gamma_n + \dot\alpha_n +
\bigg( 6H+\frac{\ddot H}{\dot H} \bigg) \alpha_n = 0 \,.
\end{equation}
The obtained ordinary differential equations are not all mutually
independent. Indeed, a straightforward calculation shows that the
identity
\begin{equation}\label{29}
\begin{array}{rl}
&\ds \left(n-1\right) \left[\left(n+2\right) a A_{n+2} -
\left(aB_n\right)_{,0} + a^3\dot H D_n \right] +            \\
&\ds \left(n+2\right) aH C_n
- 3H\left(a^3 A_n\right)_{\!,0} \equiv 0
\end{array}
\end{equation} 
holds true. It implies that Eq.\ (\ref{27b}) should be abandoned, as
it follows from Eqs.\ (\ref{25}), (\ref{26b}) and (\ref{28}). The
remaining independent equations are then readily solved. In the first
step, we consider the case $n=0$. It leads to
\begin{equation}\label{30}
\alpha_0 = \beta_0 = \frac{\ell}{a} \,,
\end{equation} 
where $\ell$ is a constant with the dimension of length. In the second
step, the remaining equations are rewritten in the form of recurrent
relations. They read
\begin{subequations}\label{31}
\begin{equation}\label{31a}
\dot\beta_n + 2\dot\gamma_n = 
-\bigg[ \dot\alpha_n + \bigg(6H + 
\frac{\ddot H}{\dot H}\bigg)\alpha_n \bigg] ,
\end{equation} 
\begin{equation}\label{31b}
\beta_{n+2} - (n+2) \gamma_{n+2} = 
\frac{a^2 H}{n+2} \bigg[ \dot\alpha_n + 
\bigg( 3H - \frac{\dot H}{H} + 
\frac{\ddot H}{\dot H} \bigg) \alpha_n \bigg] ,
\end{equation} 
\begin{equation}\label{31c}
\alpha_{n+2} = 
\frac{1}{n+1} \frac{1}{H} \left\{ \frac{a^2 H}{n+2} 
\bigg[ \dot\alpha_n + \bigg( 3H - \frac{\dot H}{H} +
\frac{\ddot H}{\dot H} \bigg) \alpha_n \bigg] \right\}_{\!\!,0}
\end{equation} 
\end{subequations}
and
\begin{equation}\label{32}
\gamma_1 = \beta_1 .
\end{equation} 
These equations determine the coefficients $\alpha_n$, $\beta_n$ and
$\gamma_n$. The only undetermined coefficient is $\alpha_1$. The {\it
constant $\ell$ and the function $\alpha_1(t)$ are the unique free
parameters} of the theory. This is because the free integration
constants of Eq.\ (\ref{31a}) turn out to be pure gauge. The proof
goes as follows. First, the power expansion
$$
\xi(r) \equiv \sum_{n=0}^{\infty}\xi_n r^n 
$$
is used to bring the residual gauge transformations Eq.\ (\ref{23}) to
the form
\begin{equation}\label{33}
\delta_0\alpha_n = 0 \,, \quad
\delta_0\beta_n = -2n\xi_n \,,\quad
\delta_0\gamma_n = -2\xi_n \,. 
\end{equation} 
Second, Eq.\ (\ref{31a}) is solved for $\beta_n+2\gamma_n$.
One obtains
\begin{equation}\label{34}
\beta_n+2\gamma_n =
-\int_0^t \bigg[ \dot\alpha_n + \bigg(6H + 
\frac{\ddot H}{\dot H}\bigg)\alpha_n \bigg] dt + c_n \,,
\end{equation} 
where $c_n$ are free integration constants. Finally, the
transformation law
$$
\delta_0 c_n = -2\left(n+2\right)\xi_n
$$
is derived by applying the gauge transformations Eq.\ (\ref{33}) to
the general solution Eq.\ (\ref{34}). It is seen that the integration
constants $c_n$ can all be gauged away, leaving us with no residual
gauge symmetry. The unique free parameters of the gauge fixed theory
are $\ell$ and $\alpha_1$.

Physical meaning of the parameters $\ell$ and $\alpha_1$ will be
addressed in the next section, where their influence on the
gravitational force is examined. Here, we anticipate that $\ell$
measures the Schwarzschild radius, or equivalently, mass of the
gravitational source, whereas $\alpha_1$ stems from the additional
inflaton degree of freedom. Indeed, $\ell = 0$ implies the absence of
$1/r$ terms in the expansion Eq.\ (\ref{24}), which leaves us with the
gravitational force without the familiar Newtonian contribution.
Nevertheless, the gravitational force does not vanish. Instead, it is
proportional to the arbitrary function $\alpha_1(t)$, which indicates
the presence of an additional degree of freedom. In the next section,
I shall examine how the choice of $\alpha_1$ influences geometry of
the point source.

Before we go on, let me derive the formula for gravitational
acceleration. One starts with the geodesic equation
$$
\frac{du^{\mu}}{ds} + \Gamma^{\mu}{}_{\nu\rho}
u^{\nu}u^{\rho} = 0 \,,
$$
where $u^{\mu} \equiv dx^{\mu}/ds$. The goal is to calculate
trajectories of stars in galaxies. For this purpose, the
non-relativistic approximation is known to work well. Thus, the
geodesic equation for the metric Eq.\ (\ref{17}) is brought to the form
\begin{equation}\label{35}
\frac{dv^{\alpha}}{dt} = 
\frac{\mu'_1}{2a^2} \frac{x^{\alpha}}{r} + {\cal O}(v) \,, 
\end{equation} 
where $v^{\alpha} \equiv dx^{\alpha}/dt$. What remains to be done is
to rewrite this equation in terms of physical distance and velocity.
In the approximation we work with, the physical distance is defined by
the background metric Eq.\ (\ref{4}). It tells us that
$$
dr_{\rm phys} = a(t) dr \,,
$$
which should be integrated out to give the global physical distance
$r_{\rm phys}$. As meaningful notion of global distance is
known to require static geometry, we shall restrict to small time
intervals in which $a(t)$ remains practically unchanged. Then, one
finds
$$
r_{\rm phys} \approx a(t_*)\,r
$$
for all $t$ in the vicinity of $t_*$. One can think of $t_*$ as the
epoch the observed galaxy lives in. For closest galaxies, it is the
present epoch $t_0$. The time $|t-t_*|$, on the other hand, is related
to how long it takes the light to travel across the galaxy. Typically,
the expansion of the Universe during this time is negligible. The
physical velocity $\vec v_{\rm phys} \equiv d\vec r_{\rm phys}/dt$ is
derived straightforwardly. One finds
$$
\vec v_{\rm phys} = a(t_*)\,\vec v \,.
$$

The time $t_*$ in the above formulas is a fixed time. It should be
emphasized, however, that $t_*$ is allowed to have different values,
depending on what specific galaxy is considered. In what follows, I
shall replace $t_*$ with more common $t$. One should only be careful
not to do this during actual calculations. The replacement $t_*\to t$
is reserved for final expressions.

Now, we are ready to rewrite Eq.\ (\ref{35}) in terms of
physical quantities. One obtains
\begin{equation}\label{36}
g = \frac{\mu'_1}{2a} \,,
\end{equation} 
where $g$ is the magnitude of the physical acceleration $\vec g \equiv
d \vec v_{\rm phys}/dt$. (The full $\vec g$ is obtained by multiplying
$g$ with $\vec r/r$.) Eq.\ (\ref{36}) holds up to ${\cal O}(v)$ terms,
which are negligible when typical galactic curves are considered. In
the next section, a specific example will be examined. The needed
$\mu$ component of the metric will be derived by solving the recurrent
relations Eq.\ (\ref{31c}). As seen from Eq.\ (\ref{36}), this is all
we need to study the problem of anomalous galactic curves.

\section{Example}\label{Sec5}

\begin{figure}[htb]
\begin{center}
\includegraphics[height=4cm]{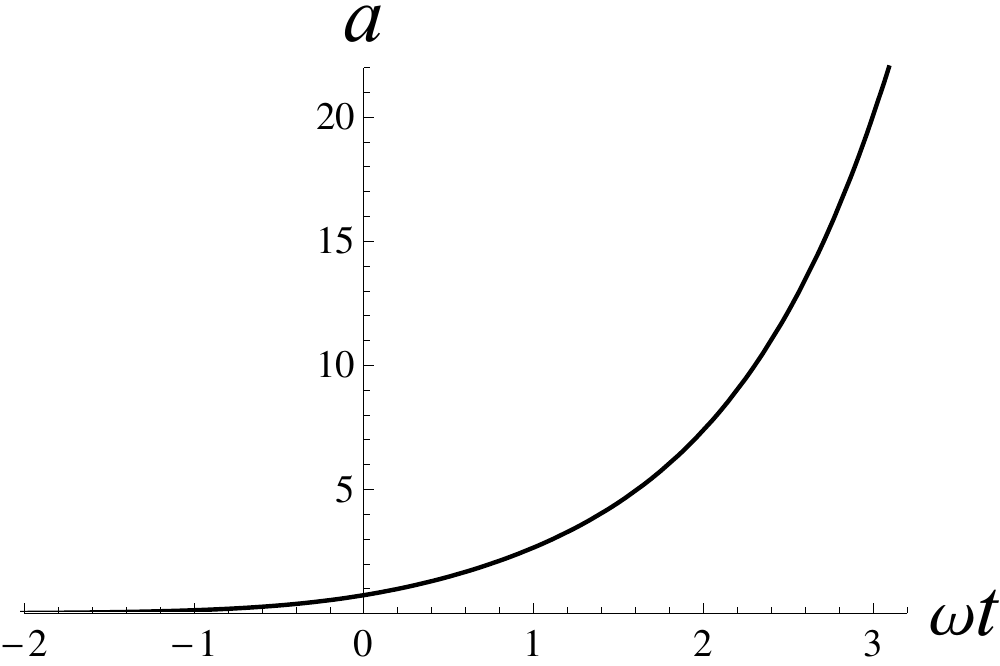}
\end{center}
\vspace*{-.5cm}
\caption{Background geometry.\label{f1} }
\end{figure}

Let me consider one specific example. For this purpose, I choose one of
the geometric sigma models defined in Sec.\ \ref{Sec2}. As these are
parametrized by their background geometries, the choice is made by
specifying the scale factor $a(t)$. In this section, I choose
\begin{equation}\label{37}
a = \ln \left[ 2\cosh(e^{\omega t}) - 1 \right] .
\end{equation} 
The graph of this scale factor is shown in Fig.\ \ref{f1}. It
represents a Universe that expands from a higher curvature de Sitter
space at $t \to - \infty$ to a lower curvature de Sitter space at $t
\to\infty$. The constant $\omega$ is a free parameter with the
dimension of mass. The scale factor Eq.\ (\ref{37}) is a solution of
the sigma model Eq.\ (\ref{7}) in which the target metric $F(\phi)$
and the potential $V(\phi)$ are defined by the replacement $t \to
\phi$ in the expressions Eq.\ (\ref{8}). As their explicit form is not
needed for the forthcoming analysis, I choose not to display them
here.

\subsection{Gravitational acceleration}\label{Sec5a}

\begin{figure}[htb]
\begin{center}
\includegraphics[height=4cm]{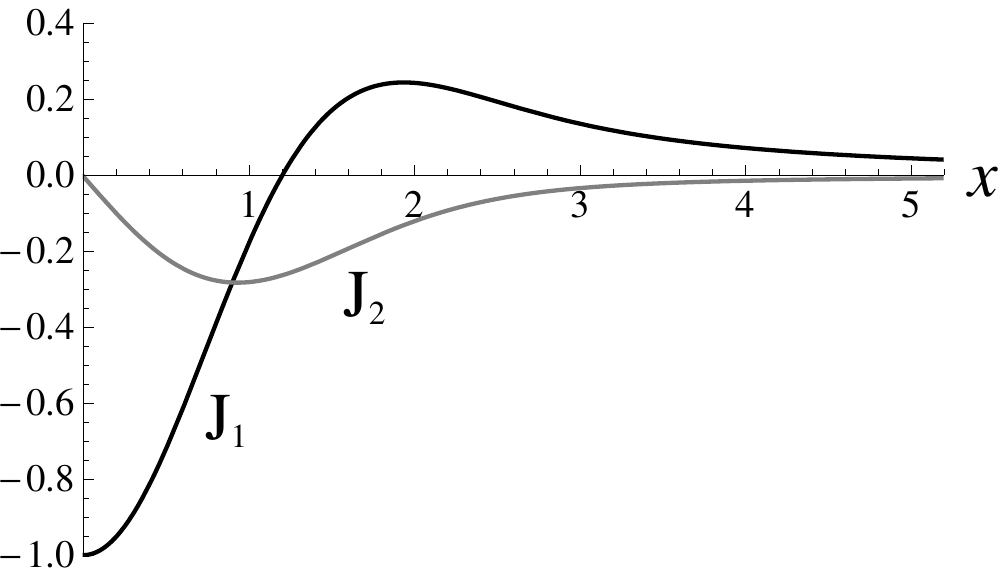}
\end{center}
\vspace*{-.5cm}
\caption{Two contributions to $\Delta g$.\label{f2} }
\end{figure}

Let me now solve the recurrent relations Eq.\ (\ref{31c}). To make it
easier, I adopt two additional simplifications. First, I restrict my
considerations to late times, where $a(t)$ is well approximated by the
exponential function. This is the late de Sitter phase of the
cosmological evolution, where our current epoch belongs to. Second,
the parameter $\alpha_1(t)$ is chosen to be
\begin{equation}\label{38}
\alpha_1 \equiv q \,,
\end{equation} 
where $q$ is unspecified dimensionless constant. With these
assumptions, the coefficients $\alpha_n$ are found to take the
following approximate values:
$$
\begin{array}{lll}
\ds\alpha_0 = \frac{\ell}{a} \,, 
&\,&\ds\alpha_1 = q \,,                                                    \\ [1.3ex]
\ds \alpha_2 \approx -\ell\,\omega^2 a^2,  
&\,&\ds \alpha_3 \approx -\frac{1}{2}q\,\omega^2 a^3,  \\ [1.3ex]
\ds\alpha_4\approx\frac{5}{12}\ell\,\omega^4 a^5,
&\,&\ds\alpha_5\approx\frac{3}{20}q\,\omega^4 a^6,    \\ [1.3ex]
\ds\alpha_6\approx -\frac{1}{9}\ell\,\omega^6 a^8,
&\,&\ds\alpha_7\approx -\frac{9}{280}q\,\omega^6 a^9 
\end{array}
$$
and so on. This pattern gives us two separate formulas for even and
odd coefficients. The even coefficients are collected in
\begin{subequations}\label{39}
\begin{equation}\label{39a}
\alpha_0 = \frac{\ell}{a} \,, \quad
\alpha_{2n} \approx \ell\,\omega^{2n} a^{3n-1}c_n 
\end{equation} 
for all $n\geq 1$. The numerical coefficients $c_n$ have the values
\begin{equation}\label{39b}
c_n \equiv \frac{(-1)^n}{(2n)!}\prod_{k=1}^{n}(3k-1) \,.
\end{equation} 
\end{subequations}
The odd coefficients are given by
\begin{subequations}\label{40}
\begin{equation}\label{40a}
\alpha_{2n+1} \approx
q\, \omega^{2n} a^{3n} d_n  
\end{equation} 
with
\begin{equation}\label{40b}
d_n \equiv (-3)^n \frac{n!}{(2n+1)!} \,.
\end{equation}
\end{subequations} 
These hold true for all $n \geq 0$. With the known $\alpha_n$
coefficients, it is straightforward to calculate the $\mu$ component
of the metric, and subsequently, the gravitational acceleration
Eq.\ (\ref{36}). The result is most conveniently expressed in terms of
\begin{equation}\label{41}
x \equiv \omega r a\sqrt{a} \,, 
\end{equation}
which serves as a dimensionless measure of spatial distance. Then, the
gravitational acceleration becomes
\begin{equation}\label{42}
g = \frac{\ell}{2} \omega^2 a
\left[-\frac{1}{x^2} + J_1(x)\right] +
q\,\omega\sqrt{a}\,J_2(x) \,,
\end{equation}
where
\begin{equation}\label{43}
J_1 \equiv \sum_{n=1}^{\infty} (2n-1)\, c_n\, x^{2n-2}, \quad
J_2 \equiv \sum_{n=1}^{\infty}n\,d_n\,x^{2n-1}. 
\end{equation}
The leading term in Eq.\ (\ref{42}) is recognized as the Newtonian
gravitational acceleration. Indeed, the latter is given by
\begin{equation}\label{44}
g_N \equiv - \frac{GM}{r^2_{\rm phys}}\,,
\end{equation}
where $G$ is the gravitational constant and $M$ denotes the source
mass. It is straightforward to verify that $g_N=-\ell \omega^2 a / 2
x^2$, once the parameter $\ell$ is identified with the Schwarzschild
radius. In terms of the source mass,
\begin{equation}\label{45}
\ell \equiv 2GM \,.
\end{equation}
Thus, the gravitational acceleration is a sum 
$$
g = g_N + \Delta g \,,
$$
where
$$
\Delta g \equiv \frac{\ell}{2}\omega^2 a\, J_1(x) +
q\,\omega\sqrt{a}\,J_2(x) \,.
$$
The first term in $\Delta g$ is a universal modification that does not
depend on the inflaton degree of freedom. It remains the same
irrespectively of the choice of the free parameter $\alpha_1(t)$. The
second, on the other hand, changes whenever the inflaton initial
conditions are changed. The graphs of $J_1$ and $J_2$ are displayed in
Fig.\ \ref{f2}. It is seen that their contribution to the
gravitational
acceleration Eq.\ (\ref{42}) can be both, positive and negative. At
small distances, they definitively increase the attractive Newtonian
force, irrespectively of the sign of the free parameter $q$. At large
distances, however, the gravitational force can become repulsive, as
expected in the expanding Universe.

\subsection{Comparison with MOND}\label{Sec5b}

In what follows, I shall examine how close to MOND the modified
gravitational force can be. Let me start with the MOND formula 
\begin{equation}\label{46}
(\Delta g)_{\rm MOND}=-\frac{\sqrt{GM g_0}}{r_{\rm phys}} \,,
\end{equation}
where $g_0$ is a constant with the dimension of acceleration
\cite{9,10,10a}. The astronomical observations indicate that the value
of $g_0$ is close to $H_0/2\pi$, where $H_0$ is the Hubble parameter
of the present epoch \cite{19,20,21,22,23,24,25,26}. In our example,
$H_0$ can be identified with the parameter $\omega$. Indeed, the late
time behavior $a \sim e^{\omega t}$ indicates that $H\sim\omega$. This
is the vacuum value of $H$, as no ordinary matter is considered. The
inclusion of ordinary matter, however, has little influence on the
measured value of the Hubble parameter. Because of this, I adopt
the approximation
\begin{equation}\label{47}
g_0 \approx \frac{\omega}{2\pi} \,.
\end{equation}
Let me now examine the equality
$$
\Delta g = (\Delta g)_{\rm MOND} \,.
$$
With the adopted approximations, it reduces to
\begin{equation}\label{48}
\sqrt{\pi\ell\omega a} = - \frac{1}{\left(J_1+p J_2\right)x} \,,
\end{equation}
where 
\begin{equation}\label{49}
p \equiv \frac{2q}{\ell\omega\sqrt{a}} \,.
\end{equation}
The fact that $t$ in $a(t)$ is fixed by the choice of the observed
galaxy makes the parameter $p$ a substitute for the free constant $q$.
In what follows, Eq.\ (\ref{48}) will be solved for a variety
of numerical values of $p$. Precisely, I search for the range of
distances for which $J_1+p J_2$ behaves as $1/x$. This is done by a
graphical method. In the first step, the graphs of the functions
$(J_1+p J_2)x$ are drawn for different values of $p$. Then, flat
portions of these curves are identified. The outcome of this analysis
is shown in Table \ref{t1}. In the first column, the value of the free
\begin{table}[htb]
\caption{Distances at which $g$ reduces to MOND. 
\label{t1}}
\vspace*{3mm}
\begin{ruledtabular}
\begin{tabular}{ccccc}
$p$ & $x$ & $\left(J_1+p J_2\right)x$ &
$\ell\omega a$ & $\ds\frac{q}{\sqrt{\ell\omega}}$   \\ 
[1.3ex] \hline
$1$ & $(0.62\,,\,0.86)$ & $-0.529$ &
$1.14$  & 0.53     \\
$3$  &  $(0.82\,,\,1.12)$ & $-0.998$ &
$3.2\times 10^{-1}$  & 0.85   \\
$9$     &  $(1.03\,,\,1.38)$ & $-2.768$ &
$4.2\times 10^{-2}$  & 0.91   \\
$27$   &  $(1.12\,,\,1.50)$ & $-8.393$ &
$4.5\times 10^{-3}$  & 0.91   \\
$81$ & $(1.15\,,\,1.55)$ & $-25.36$ &
$4.9\times 10^{-4}$  & 0.91     \\
$243$  &  $(1.16\,,\,1.55)$ & $-76.35$ &
$5.5\times 10^{-5}$  & 0.89   \\
$729$     &  $(1.17\,,\,1.55)$ & $-229.5$ &
$6.0\times 10^{-6}$  & 0.90  \\
$2187$   &  $(1.16\,,\,1.56)$ & $-687.8$ &
$6.7\times 10^{-7}$  & 0.89   \\
\end{tabular}
\end{ruledtabular}
\end{table}
parameter $p$ defines which function $(J_1+p J_2)x$ is examined. As it
turns out, each of these functions can be approximated by a constant
in some interval of distances. These intervals are shown in the second
column. The corresponding constant values of the functions $(J_1+p
J_2)x$ are displayed in the third column. The last two columns are
reserved for quantities that are straightforwardly derived from Eqs.\
(\ref{48}) and (\ref{49}). The error that appears when portions of
curves are approximated by constants is kept lower than $2\%$.
Considering the typical precision of astronomical measurements, this
is a very good approximation.

In what follows, I shall inspect the collected data more closely. Let
me start with the last column of Table \ref{t1}. It tells us that the
expression $q/\sqrt{\ell\omega}$ is practically independent of $\ell$.
Precisely,
\begin{equation}\label{50}
q \approx 0.9 \sqrt{\ell\omega} 
\end{equation}
for all $\ell$ that satisfy $\ell\omega a < 0.1$. Only then, the
gravitational acceleration Eq.\ (\ref{42}) has regions with MOND-like
behavior. The range of $\ell$ for which $q(\ell)$ has the form
Eq.\ (\ref{50}) depends on time. Indeed, the time $t$ in $a=a(t)$ varies
from one observed galaxy to another. For closest galaxies, $t$ takes
the present value $t_0$. It is determined by the current values of the
Hubble and deceleration parameters, and so is the parameter $\omega$.
Instead of repeating these well known calculations, let me summarize
the results. First, the parameter $\omega$ is identified with the
current value of the Hubble parameter. This is because the late time
approximation I use in this paper implies the exponential law $a\sim
e^{\omega t}$. A rough estimation is that
$$
\omega \approx 10^{-10}\ {\rm yr}^{-1} .  
$$
Second, the present time $t_0$, apart from belonging to the late de
Sitter phase of the cosmological evolution, is freely chosen. This is
possible because the relevant formulas have one free integration
constant. This integration constant is related to the freedom of
defining the origin of time. (The known result $t_0 \approx 13 \ {\rm
Gyr}$ is obtained when the origin $t=0$ is associated with the initial
singularity.) In accordance with the assumed late time approximation
($\omega t \gg 1$), the present time $t_0$ is chosen to be
$$
\omega t_0 = 23 \,.
$$
It implies $a_0\approx 10^{10}$, so that the inequality $\ell\omega a
< 0.1$ turns into
\begin{equation}\label{51}
\ell < 10^{-1} \ {\rm ly}
\end{equation}
at present time. The Schwarzschild radius of $10^{-1}$ ly is known to
correspond to the largest galaxies in the observed Universe. Thus, the
restriction Eq.\ (\ref{51}) does not rule out any of the
observationally interesting astronomical objects in our vicinity.

\begin{figure}[htb]
\begin{center}
\includegraphics[height=4cm]{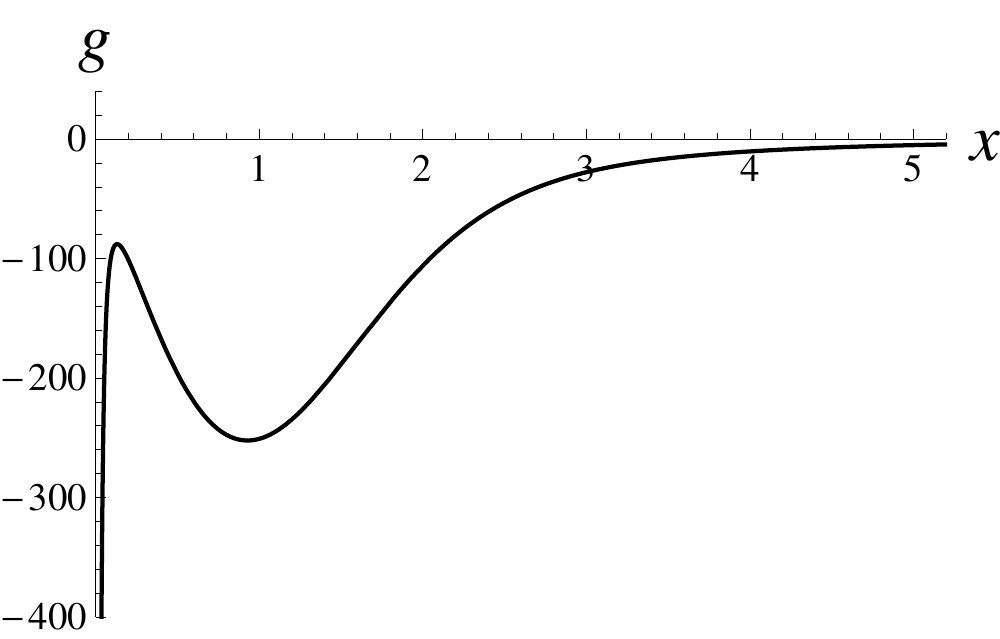}
\end{center}
\vspace*{-.5cm}
\caption{Gravitational force of Milky Way black hole.\label{f3} }
\end{figure}

Further inspection of data in Table \ref{t1} reveals that the obtained
MOND behavior is not as universal as the original MOND suggests.
Indeed, it is only a small interval of distances far from the point
source where the modified gravitational law Eq.\ (\ref{42}) reduces to
MOND. These distances are shown in the second column of Table
\ref{t1}. For not too large source masses, $x \in (1.15, 1.55)$, which
corresponds to
\begin{equation}\label{52}
1.15 \times 10^5\ {\rm ly} < r_{\rm phys} < 
1.55 \times 10^5\ {\rm ly}. 
\end{equation}
For comparison, the largest galaxies in the observable Universe are
about $10^5$ ly in diameter. Thus, the stars in small galaxies
($10^3\,$--$\,10^4$ ly) {\it do not feel MOND regime}. As an
illustration, the gravitational acceleration Eq.\ (\ref{42}) for $\ell
\approx 10^{-6}$ ly is depicted in Fig.\ \ref{f3}.
This value of $\ell$ corresponds to the Schwarzschild radius of the
black hole in the center of Milky Way. (The unit value of $g$ in the
graph is $10^{-16}$ ${\rm yr}^{-1}$.) It is seen that modified
gravitational acceleration $g$ differs quite a lot from the Newtonian
acceleration $g_N$. This is certainly true in the interval
$x\in(0.5,10)$, where the magnitude of $g$ considerably exceeds that
of $g_N$. It should be noted, however, that this modification has
little to do with MOND. Indeed, MOND behavior is found only in the
small interval $x\in(1.15,1.55)$. For comparison, the diameter of
Milky Way is $x \sim 1$. The Newtonian gravitational force is
recovered in the limit $x \to 0$. As for $x \gg 1$, the gravitational
acceleration becomes positive. It corresponds to the repulsive
gravitational force that drives the expansion of the Universe.

In conclusion, I demonstrated that scalar-tensor theories may indeed
have solutions that mimic MOND. The problem is that these solutions do
not cover all the distances that the original MOND suggests. Of
course, this is not a generic feature of the class of models
considered in this paper. Hopefully, a different geometric sigma model
could have a better behaving solution. Another possibility to get
closer to MOND is to make a different choice of the free parameter
$\alpha_1(t)$. Instead of the simplest choice $\alpha_1 = {\rm
const.}$, one may try $\alpha_1$ with a nontrivial time dependence.
The problem is that one has no clue how to choose $\alpha_1(t)$ to
obtain the desired result. In the next section, I shall do a
nonperturbative analysis of the field equations, with the idea to
replace the free parameter $\alpha_1(t)$ with some standard initial
conditions. Some of these initial conditions will be shown to allow for
exact MOND at a fixed moment of time.

\section{Nonperturbative approach}\label{Sec6}

Let me consider Eqs.\ (\ref{22}) once again. It is seen that Eqs.\
(\ref{22a}) and (\ref{22b}) can be expressed in terms of $\mu_1$ and
$\nu_1-\left(r\rho_1\right)'$, only. Thus, I define the new variables
\begin{equation}\label{53}
u \equiv rH\mu_1 \,, \qquad
v \equiv r\left[\nu_1-\left(r\rho_1\right)'\right] .
\end{equation}
In terms of these new variables, Eqs.\ (\ref{22a}) and (\ref{22b})
become
\begin{equation}\label{54}
\dot v = r u' - u \,, \qquad
\dot u = \frac{1}{r a^2} v' - Q u \,,
\end{equation}
where
\begin{equation}\label{55}
Q \equiv 3H - 2\frac{\dot H}{H} + \frac{\ddot H}{\dot H} \,.
\end{equation}
This is a system of two first order partial differential equations,
solved for $\dot u$ and $\dot v$. Hence, the solutions are naturally
parametrized by the initial conditions $u(t_0,r)=u_0(r)$ and
$v(t_0,r)=v_0(r)$. The problem is that $u_0(r)$ and $v_0(r)$ cannot be
arbitrarily chosen. This is because the variables $u$ and $v$ are
subject to more than just two equations (\ref{54}). As a consequence,
their initial values are partially restricted. It is seen from the
perturbative result of Sec.\ \ref{Sec4} that only odd coefficients in
the power expansion Eq.\ (\ref{24}) carry some arbitrariness. Indeed,
these are proportional to the arbitrary function $\alpha_1(t)$ and its
derivatives, whereas even coefficients are not. In what follows, I
shall consider strictly odd functions of $r$. Then, the allowed
initial conditions have the form
$$
u_0 = \sum_{n=0}^{\infty}a_{n} r^{2n+1} , \quad
v_0 = \sum_{n=0}^{\infty}b_{n} r^{2n+3},
$$
where $a_{n}$ and $b_{n}$ take arbitrary values. (The absence of
linear term in the power expansion of $v_0$ is a consequence of
Eq.\ (\ref{32}).) The arbitrariness of $a_{n}$ and $b_{n}$, on the
other hand, implies some freedom in choosing the initial values of $u$
and $\dot u$. Precisely, the {\it odd parts of the functions
$u(t_0,r)$ and $\dot u(t_0,r)$ can be freely chosen}. This can be
utilized in the second order differential equation for the variable
$u$, which is easily derived from Eq.\ (\ref{54}). One finds
\begin{equation}\label{56}
\ddot u + \big(Q+2H\big)\dot u + \big(\dot Q+2HQ\big) u =
\frac{1}{a^2} u'' .
\end{equation}
The partial differential equation (\ref{56}) is too complicated to be
solved analytically. Numerical calculations, on the other hand,
require additional adaptations. First, all the variables and
coefficients should be made dimensionless. The easiest way to achieve
this is to fix the system of units. Let me choose
\begin{equation}\label{57}
\omega =1 \,.
\end{equation}
In this system of units, the unit distance is the Hubble distance $r_H
\equiv 1/\omega$. Its approximate value is $10^{10}$ ly. In accordance
with the adopted natural units $c=\hbar=0$, the corresponding
approximate value of the unit time is $10^{10}$ years. Second, the
initial conditions should be specified. In this section, I choose the
present epoch $t_0$ to play the role of the initial time. The ideal
choice of the initial value of $u$ would be
\begin{equation}\label{58}
u_0 = \frac{\ell}{a_0} - 
\sqrt{\frac{\ell}{\pi}}\,r\big(\ln r + {\rm const}\big) ,
\end{equation}
as it implies exact MOND at $t=t_0$. Unfortunately, the choice
Eq.\ (\ref{58}) is not possible. This is because the even part of
Eq.\ (\ref{58}) does not agree with the perturbative result of
Sec.\ \ref{Sec4}. The odd part of $u_0$, however, can be freely
chosen. In accordance with this, I adopt
\begin{equation}\label{59}
u_0 = r\ln r \,, \quad \dot u_0 = 0 \,.
\end{equation}
The choice Eq.\ (\ref{59}) differs from the ideal choice Eq.\
(\ref{58}) in two respects. First, the Newtonian term $\ell/a_0$ in
Eq.\ (\ref{58}) is missing in Eq.\ (\ref{59}). However, this term has
negligible influence on the flat galactic curves we are interested in.
Indeed, if we restrict to regions far from the gravitational source,
the initial conditions Eq.\ (\ref{59}) are practically indistinguishable
from Eq.\ (\ref{58}). The second difference is that $u_0$ of Eq.\
(\ref{59}) lacks the constant $-\sqrt{\ell/\pi}$. This constant,
however, has been deliberately omitted for simplicity. It can be
restored later because the general solution of Eq.\ (\ref{56}) is
determined only up to a multiplicative constant. Finally, let me
explain why the function $r\ln r$ is considered odd. The reason is the
regularization scheme that I use to replace the singular function $\ln
r$ with everywhere regular, even function $\frac 12
\ln{(r^2+\epsilon^2)}$. At large distances, or small values of
$\epsilon$, the latter is well approximated by $\ln r$.

\begin{figure}[htb]
\begin{center}
\includegraphics[height=4cm]{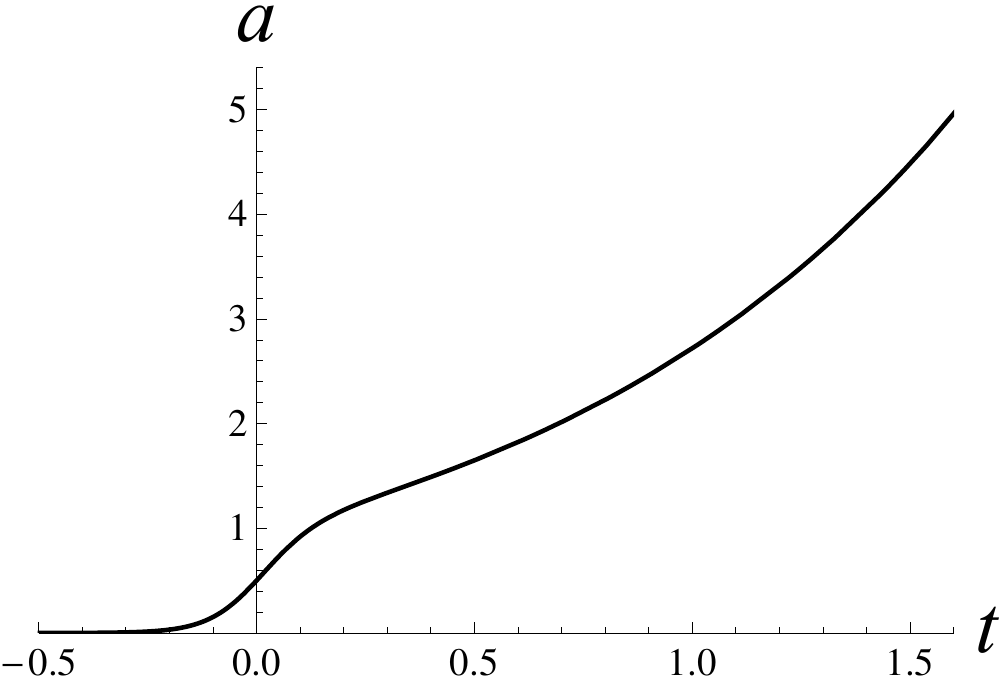}
\end{center}
\vspace*{-.5cm}
\caption{Another background geometry.\label{f4} }
\end{figure}

To recapitulate, I have demonstrated that {\it it is always possible
to have MOND at a fixed moment of time}. It is achieved by imposing
proper initial conditions, such as Eq.\ (\ref{59}). What one should
check is how this result evolves with time. Let me solve
Eq.\ (\ref{56}) numerically. In the first step, one specifies the
support of the function $u(t,r)$. This is done as follows. First, one
determines the time interval in which physically relevant observations
are made. It is well known that flat galactic curves are detected in
most of the surrounding galaxies, including very distant ones. I am
talking about distances up to $10^8$ ly. Thus, the time coordinate $t$
should go at least $10^8$ years to the past. At the same time, the
closest of the observed flat galactic curves is about $10^5$ ly far
from us. In the adopted system of units, this implies
\begin{subequations}\label{60}
\begin{equation}\label{60a}
-10^{-2} < t - t_0 < -10^{-5} \,.
\end{equation}
The range of $r$ coordinate is related to the radii of the observed
galactic curves. In terms of physical distance, this range is defined
by $r_{\rm phys} \in (10^2,10^5)$ ly, which leads to
\begin{equation}\label{60b}
10^{-8} < r a_0 < 10^{-5}
\end{equation}
\end{subequations}
in the system of units $\omega = 1$. The exact numerical values are
obtained when the present time $t_0$, and the scale factor $a(t)$ are
specified. As an example, I shall consider the scale factor
\begin{equation}\label{61}
a = \frac{e^{t}}{1+e^{-16 t}} \,.
\end{equation}
Its graph is shown in Fig.\ \ref{f4}. The present time $t_0$ must
belong to the late de-Sitter phase of the given geometry. This leads
me to choose
$$
t_0 = 1 \,.  
$$
The support of the function $u(t,r)$ is then defined by
$$
t \in (0.9\,, 1) \,, \quad
r \in (10^{-8}, 10^{-5}) \,.
$$
This region of $t$--$r$ plane does not exactly coincide
with Eq.\ (\ref{60}), but is more suitable for numerical calculations.

The solution of Eq.\ (\ref{56}) with the initial conditions (\ref{59})
is easily seen to have the structure
$$
u = f_0\, r \ln r + \sum_{n=1}^{\infty} 
\frac{f_n}{r^{2n-1}} \,,
$$
where $f_n = f_n(t)$ for all $n \geq 0$. As a consequence, the partial
differential equation (\ref{56}) is rewritten as the system of
ordinary differential equations
$$
\ddot f_0 + \big(Q+2H\big) \dot f_0 + 
\big(\dot Q+2QH\big) f_0 = 0 \,,
$$
$$
\ddot f_1 + \big(Q+2H\big) \dot f_1 + 
\big(\dot Q+2QH\big) f_1 = \frac{1}{a^2} f_0 \,,
$$
$$
\begin{array}{rc}
&\ds \ddot f_n + \big(Q+2H\big) \dot f_n + 
\big(\dot Q+2QH\big) f_n =                          \\ [0.6ex]
&\ds \frac{(2n-3)(2n-2)}{a^2} f_{n-1} 
\end{array}
$$
for all $n\geq 2$. The corresponding initial conditions are
$$
\begin{array}{ll}
& f_0(1) = 1 \,,\quad \dot f_0(1) = 0 \,,      \\
& f_n(1) = 0 \,,\quad \dot f_n(1) = 0 
\end{array}
$$
for all $n \geq 1$. The numerical solution for the first several
functions $f_n(t)$ is obtained straightforwardly. As it turns out,
$f_n(t)$ are well approximated by
$$
f_n \approx \kappa_n \left(1-t\right)^{2n}
$$
in the interval $t\in (0.9\,, 1)$. The constants $\kappa_n$ take
values
$$
\begin{array}{ll}
\kappa_0 = 1 \,,                           &
\kappa_1 = 6\times 10^{-2} ,      \\
\kappa_2 = 1.4\times 10^{-3},\ \ &                
\kappa_3 = 7.5\times 10^{-5} ,   \\
\kappa_4 = 3.7\times 10^{-6} ,   &                 
\kappa_5 = 5\times 10^{-7} ,      \\
\kappa_6 = 5\times 10^{-8} ,      &                 
\kappa_7 = 5\times 10^{-9} , \ \dots \,.     
\end{array}
$$
With these results, the function $u(t,r)$ can be rewritten as
$$
u \approx r\ln r + \frac{1-t}{100} \sum_{n=0}^{\infty}
\frac{\bar\kappa_n}{(2n+1)!} 
\left(\frac{1-t}{r}\right)^{2n+1} ,
$$
where $\bar\kappa_n$ take values
$$
\bar\kappa_n = 6,\,1,\,1,\,2,\,20,\,200,\,3000,\,\dots \,.
$$
Finally, one can calculate the gravitational acceleration $g$. As seen
from Eqs.\ (\ref{36}) and (\ref{53}), the gravitational acceleration
is proportional to $(u/r)'$, so that one obtains
\begin{equation}\label{62}
g \propto \frac{1}{r} + \frac{t-1}{(10r)^2}
\sum_{n=0}^{\infty} \frac{\tilde\kappa_n}{(2n+1)!} 
\left(\frac{1-t}{r}\right)^{2n+1} ,
\end{equation}
where $\tilde\kappa_n$ take values
$$
\tilde\kappa_n = 12,\,4,\,6,\,16\,,200,\,2400,\,42000,\,\dots\ . 
$$
To estimate the value of $g$, note that $\tilde\kappa_n > 1$ implies
$$
\sum_{n=0}^{\infty} \frac{\tilde\kappa_n}{(2n+1)!} 
\left(\frac{1-t}{r}\right)^{2n+1}\!\! >\ \sinh\frac{1-t}{r}\,.
$$
It can be checked that
$$
\frac{1-t}{r} > 10
$$
for all the observed galaxies in our neighborhood. Owing to this, 
$$
g\, \propto\, \frac {1}{r}\, F \bigg(\frac{1-t}{r}\bigg) ,
$$
where
$$
F\bigg(\frac{1-t}{r}\bigg) > \frac{1-t}{r} \sinh \frac{1-t}{r} \,. 
$$
It is seen that the gravitational law rapidly changes with time. In
fact, the change is so rapid that the earlier found possibility to
have exact MOND at the initial moment of time is completely
disqualified. Indeed, even the nearest galaxies are time shifted with
respect to the present time. If MOND held true in our Milky Way, it
would definitely be lost in the surrounding galaxies. In fact, a quick
look at Eq.\ (\ref{62}) tells us that MOND can be found only in the
region $\ds\frac{1-t}{r} \sim 1$. This rules out all but our own
galaxy.

\section{Concluding remarks}\label{Sec7}

I have shown in this paper that geometric sigma models with one scalar
field can indeed have MOND-like solutions. Unfortunately, the range of
distances where the gravitational law reduces to MOND turns out to be
much smaller than needed. This is an outcome of the example considered
in Sec.\ \ref{Sec5}. Of course, one can repeat the analysis with
another geometric sigma model, or with another choice of the free
parameter $\alpha_1(t)$. While this may result in a better behaved
gravitational law, it is not clear how exactly one can achieve it.
Indeed, one can not possibly know which $\alpha_1(t)$ leads to the
desired result.

Further development has been made in Sec.\ \ref{Sec6}, where
$\alpha_1(t)$ has been replaced with some standard initial conditions.
Precisely, I have demonstrated that initial conditions for even part
of the metric component $g_{00}$ can be freely chosen. As a
consequence, MOND behavior can always be achieved at a particular
instance of time. Unfortunately, its time evolution turns out to be so
rapid that exact MOND is quickly lost. In particular, the desired MOND
behavior is preserved only in our own galaxy.

To summarize, the class of geometric sigma models with one scalar
field can hardly be responsible for the appearance of MOND. Even if it
can, there remains an important question of how fine tuned initial
conditions can be avoided. As I have already explained, the
possibility to obtain MOND relies a great deal on the extra scalar
degree of freedom. This is because, unlike vector and tensor modes,
the scalar mode is not frozen by the spherically symmetric ansatz.
Obviously, this helps the search for MOND, but it leaves us with the
problem of fine tuned initial conditions. In this respect, I believe
that scalar-vector-tensor theories are worthy of investigation. A nice
class of such models is the class of geometric sigma models with four
scalar fields \cite{18}. The particle spectrum of these models have been
proven to consist of $2$ scalar, $2$ vector and $2$ tensor degrees of
freedom, rather than $2$ tensor and $4$ scalar ones. If vector degrees
of freedom happen to be responsible for MOND, no fine tuning problem
would be left behind. The analysis of these models will be presented
elsewhere.

\begin{acknowledgments}
This work is supported by the Serbian Ministry of Education, Science
and Technological Development, under Contract No. $171031$.
\end{acknowledgments}

\end{document}